# Synthesis and magnetic properties of $Mn_{12}$-based single molecular magnets with benzene and pentafluorobenzene carboxylate ligands

V. S. Zagaynova[1], T. L. Makarova[1,2], N, G. Spitsina[3], D. W. Boukhvalov[4]

[1] *Department of Experimental Physics, Umea University, 90187 Umeå, Sweden*

[2] *Ioffe PTI RAS, 194021 St. Petersburg, Russia*

[3] *Institute of Problems of Chemical Physics RAS, 142432 Chernogolovka, Moscow Region, Russia*

[4] *Computational Materials Science Center, National Institute for Materials Science, 1-2-1 Sengen, Tsukuba, Ibaraki 305-0047, Japan*

E-mail: valeria.zagainova@physics.umu.se

## Abstract

*We report on the synthesis and magnetic properties of $Mn_{12}$-based single molecular magnets (SMMs) with benzene and pentafluorobenzene carboxylate ligands. The changes in ligand structure are shown to have a decisive effect on magnetic properties of produced complexes. The compound with benzene demonstrates unusual magnetic behaviour, namely, temperature dependencies of magnetization taken under the zero field cooled and field cooled conditions are split below 10 K and furthermore remnant magnetization and coercive force remain nonzero in this temperature range. In contrast, compound with pentafluorobenzene displays the customary signatures of a blocking temperature at 3K. The effect of ligand substitution was theoretically studied within local density approximation taking into account on-site Coloumb repulsion. Calculation results confirm that the electronic structure and the magnetic exchange interactions between different Mn atoms strongly depend on the type of ligands.*

# 1. INTRODUCTION

First SMM of formula $[Mn_{12}O_{12}(CH_3COO)_{16}(H_2O)_4] \cdot 2CH_3COOH \cdot 4H_2O$ was discovered in 1980 [1]. Since that time, family of Mn-based molecular magnets has been expanded to a great number of derivatives, but $Mn_{12}$-based molecular magnets remain the best known and the most promising. The synthesis of $Mn_{12}$-acetate is rather simple, and huge uniaxial magnetic anisotropy makes $Mn_{12}$-based magnet a good model for studying SMM quantum effects. $Mn_{12}$-acetate and related SMMs are promising candidates for high-density magnetic storage devices, spintronics components, wireless communiucations and biological applications [2-5]. However, certain aspects of their fundamental properties are still not completely understood. The crystal structure of $Mn_{12}$ clusters core contains four central $Mn^{4+}$ ions with spin 3/2 surrounded by eight $Mn^{3+}$ ions with spin 2 connected by the bridging oxygen ligands. Antiferromagnetic exchange interactions between $Mn^{4+}$ and $Mn^{3+}$ ions lead to the formation of a ferrimagnetic structure with total ground spin S = 10. The geometry of the core remains almost undisturbed by different carboxylate ligands; distances and angles between atoms are changed by less than 2%, but the electronic structure of the molecule depends on the ligands which determine its magnetic properties. To draw out the correlation between the nature of the carboxylate ligand and the magnetic behavior of the corresponding $Mn_{12}$ complex, we have studied several $Mn_{12}$ complexes differing in the nature of the peripheral carboxylate ligands. In this paper, we report on the synthesis and comparative analysis of magnetic properties for $Mn_{12}$-based SMMs with benzene $[Mn_{12}O_{12}(C_6H_5COO)_{16}(H_2O)_4]$ and pentafluorobenzene $[Mn_{12}O_{12}(C_6F_5COO)_{16}(H_2O)_4]$ ligands.

## 2. EXPERIMENTAL DETAILS

### 2.1 Synthesis

All manipulations were carried out under aerobic conditions at room temperature using commercial grade chemicals and solvents. The reaction products were characterized by elemental analysis and Raman spectroscopy.

**[Mn$_{12}$O$_{12}$(C$_6$F$_5$COO)$_{16}$(H$_2$O)$_4$] (1).** Neutral [Mn$_{12}$O$_{12}$(O$_2$CC$_6$F$_5$)$_{16}$(H$_2$O)$_4$] (**1**) was prepared using the ligand substitution method, as described in [6]. Elemental analysis: Calcd (Found) for **1** (C$_{112}$H$_8$O$_{48}$F$_{80}$Mn$_{12}$): C, 31.2 (30.8); H, 0.19 (0.3).

**[Mn$_{12}$O$_{12}$(C$_6$H$_5$COO)$_{16}$(H$_2$O)$_4$] (2).** One-pot method of the preparation of **2** was used [7]. Solid KMnO$_4$ was added in to solution of benzoic acid in MeOH and stirred for 4 h. In this period the KMnO$_4$ slowly dissolved and dark brown precipitate was collected by filtration and recrystallized from the mixture CH$_2$Cl$_2$/hexane to obtain black crystals of **2**. The identity of the product was confirmed by elemental analysis and IR spectral comparison with an authentic sample. Elemental analysis: Calcd (Found) for **2** (C$_{112}$H$_{88}$O$_{48}$Mn$_{12}$): C, 47.0 (46.0); H, 3.1 (3.0).

## 2.2 Physical measurements

DC magnetic measurements were performed at a Quantum Design SQUID magnetometer (MPMS-XL-1) in -1T – 1T magnetic fields. The dc magnetic susceptibility data were collected in the 1.76 – 300 K range; temperature depemdensies were taken in a 0.01 T magnetic field. The Raman spectra were recorded using a Raman spectrometer Renishaw 100 at wavelength 514 nm.

## 2.3. Theoretical method

Calculation of electronic structure and exchange interaction of Mn$_{12}$ compounds were carried out within local density approximation (LDA) realized within computer code LMTO-ASA Stuttgart-47 [8] with taking into account on-site Coloumb repulsion (LDA+U) [9], while the effects of the spin-orbit interaction were neglected. The account of the local Coulomb interaction is crucial for an adequate description of the transition metal-oxide systems in general, and for Mn$_{12}$-based SMM in particular. The value of the Coulomb parameter U for Mn$_{12}$ is 4 eV as has been determined earlier based on experimental and theoretical studies [10]. Other technical details of computational procedure are the same as in Ref. 10. For the calculation of electronic structure, experimental crystal structures of Mn$_{12}$O$_{12}$(RCOO)$_{16}$(H$_2$O)$_4$ were used, for R = CH$_3$ [1], C$_6$H$_5$ [11], C$_6$F$_5$ [12].

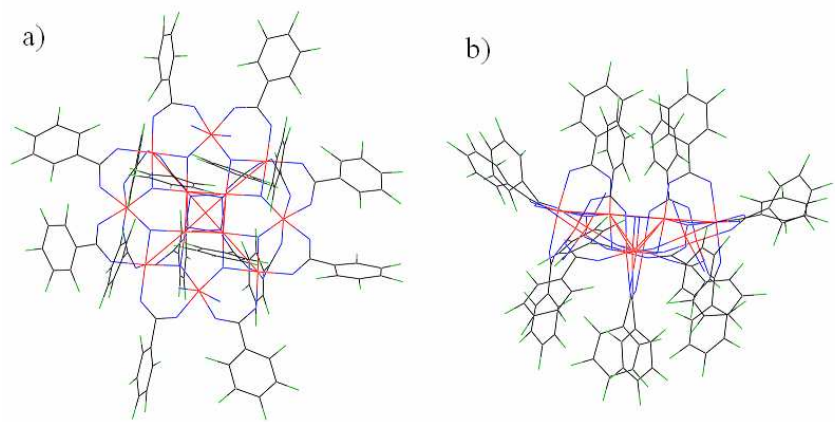

**Fig. 1.** (Color online) Atomic structure of $Mn_{12}O_{12}(C_6F_5COO)_{16}(H_2O)_4$ molecule. Top (a) and side (b) view. Manganese atoms are shown by red, oxygen by blue, carbon by black and fluorine by green. Hydrogen atoms of water and molecules of solvents are omitted for clarity.

## 3. RESULTS AND DISCUSSION

To characterize the obtained compounds, Raman spectra were recorded. In Fig. 2 the spectra of samples **1** and **2** are shown in comparison. In sample's **1** spectrum the modes at 530 cm$^{-1}$, 597 cm$^{-1}$, 650 cm$^{-1}$, 692 cm$^{-1}$ assigned to Mn-O vibrations, and the broad peak around 1420 cm$^{-1}$ assigned to Mn-acetate stretching mode are well resolved. In sample **2** the modes which are associated with the vibrations of the magnetic core of the molecule are noticeably supressed. Instead, these samples demonstrate enhanced modes of carboxyl ligands: the bands around 1606 cm$^{-1}$ and 3075 cm$^{-1}$ are assigned to the vibrations of carboxylate (OCO)$_{asym}$ and OH (coordinated $H_2O$) groups, respectively, which points out the different arrangements of bound $H_2O$ and carboxylic ligands for these structures and allows one to expect different magnetic properties.

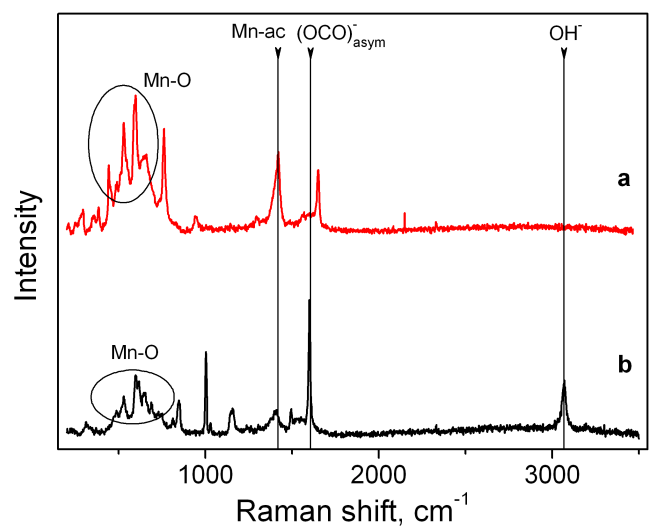

**Fig. 2.** (Color online) The Raman spectra of complexes **1** (a) and **2** (b).

Fig. 3 presents the comparison of magnetization data for complexes **1** and **2**. We note that the field-cooled (FC) and zero-field-cooled (ZFC) curves of sample **1** show magnetic behavior typical for $Mn_{12}$-based SMM complexes, and a blocking temperature $T_B$ is ~3 K.

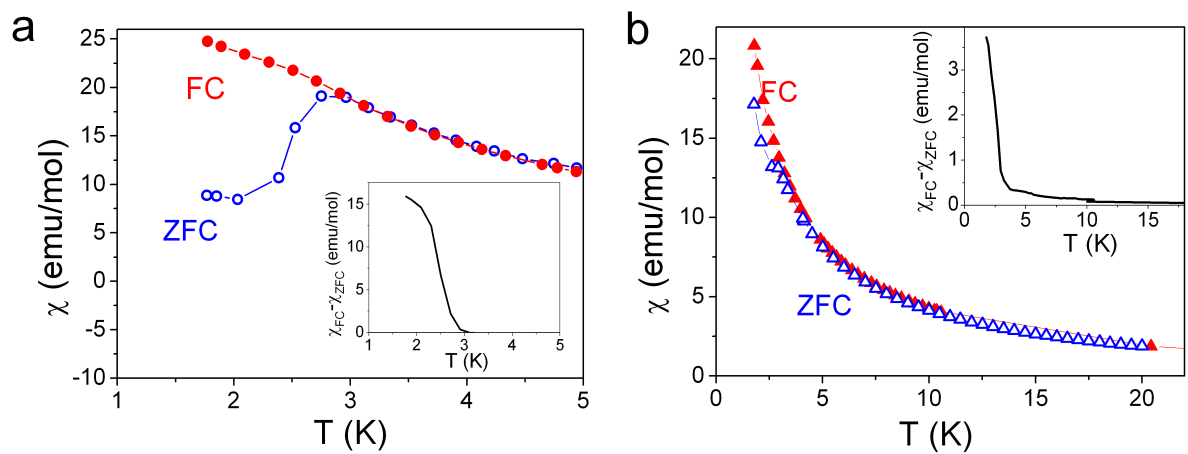

**Fig. 3.** (Color online) The FC and ZFC magnetization curves for complex **1** (a) and **2** (b); the insets show the corresponding residuals between FC and ZFC curves.

By contrast, the ZFC –FC difference complex 2 stays split well above the usual for Mn12 blocking temperature TB~3.5 K. The insets show the corresponding residuals between FC and ZFC curves. There is a clear and measurable difference between the curves at least at 15 K for complex 2: the changes in magnetic moment are 3 orders of magnitude higher than the sensitivity of the device. Similar ZFC-FC differences for Mn₁₂-acetate were described earlier in [13] and were interpreted as an increased of a blocking temperature due to the enhanced magnetic anisotropy which was in turn assigned to the molecule electronic shell modification due to the specific treatment. Another possible explanation can relate to a mixed bulk-magnet and single-molecule magnet behaviours described in Ref. [14].

The dc magnetic susceptibility data of the complexes measured in a 0.1 T field and 1.75 – 300 K temperature range are shown on Fig. 4. The curves of complex **2** on decreasing of temperature goes through broad minimum indicating ferrimagnetic ordering, and then the $\chi_M T$ value slowly increases to maximum of 46.4 ($\mu_{eff}$ = 19.3) emu·K·mol$^{-1}$ at 8 K, and afterwards decreases at lower temperatures due to Zeeman effects from the applied field or intercluster interactions. For sample **1** the minimum is not so prominent, and maximum corresponds to effective moment $\mu_{eff}$ = 21 which is consistent with the theoretical spin-only value for S=10.

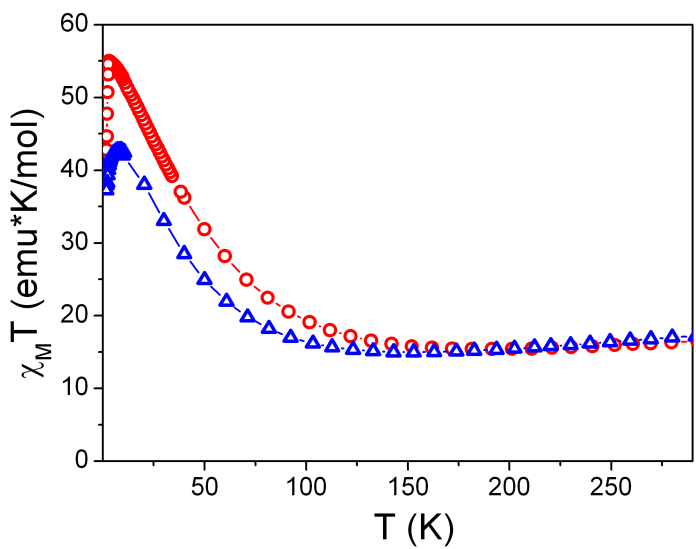

**Fig. 4.** (Color online) Temperature dependences of $\chi_M T$ for complexes **1** (circles) and **2** (triangles).

The hysteresis loops of all complexes with removed diamagnetic background from the holder, taken at various temperatures are presented on Fig. 5. These data confirm the observation obtained from the FC-ZFC curves. Whereas the complex **1** (Fig. 5a) displays usual behaviour and magnetic M(H) isotherms measured above the blocking temperature are linear, the preserved nonlinearity for complex **2** has been observed at higher temperatures (Fig. 5b). Moreover, hysteresis loops stay open at temperatures considerably higher than 3 K (Fig. 5c). Both remnant magnetization (Mr) and coercive force (Hc) gradually decrease with temperature, Mr dependence being nearly linear versus reciprocal temperature. We believe that the observed effects can be due to the disruption of zero-dimensionality of SMMs. In the studied samples the Mn12 molecules are not completely isolated by small solvent molecules, but instead form a three-dimensional system through π-π stacking between the aromatic rings of ligands. A necessary requirement to obtain SMM behaviour, namely that the intercomplex magnetic interactions should remain significantly smaller than intracomplex ones seems to be broken in complex 2, and thus the dependencies shown in Fig. 5c can be interpreted as the SMM behaviour which competes or coexists with the features of bulk magnets [14]. Insets in Figs. 5a and 5b demonstrate that neither of investigated complexes displays the steps on the hysteresis curves originating from quantum transitions (QTs), though the derivatives dM/dH allow distinguishing the QT positions in complex **2** whereas in complex **1** QTs are totally absent.

A scheme of intermolecular exchange interactions and the results of calculations of the values of exchange interactions are presented on Fig. 6. As we can see, substitution of ligands from $CH_3$ to $C_6H_5$ provide significant changes of exchange interactions that corresponds to experimentally observed changes of several magnetic properties of $Mn_{12}$ [11], and is in line with our observations (Figs. 3 and 5). The changes in electronic structure and magnetic behavior were recently discussed for another manganese based SMM [15]. Whereas in the classical $Mn_{12}$-acetate SMM all exchange couplings are antiferromagnetic (Fig. 6a), the sign of exchange integral inverts for the central and peripheral Mn atoms in case of benzene ligand (Fig. 6b).

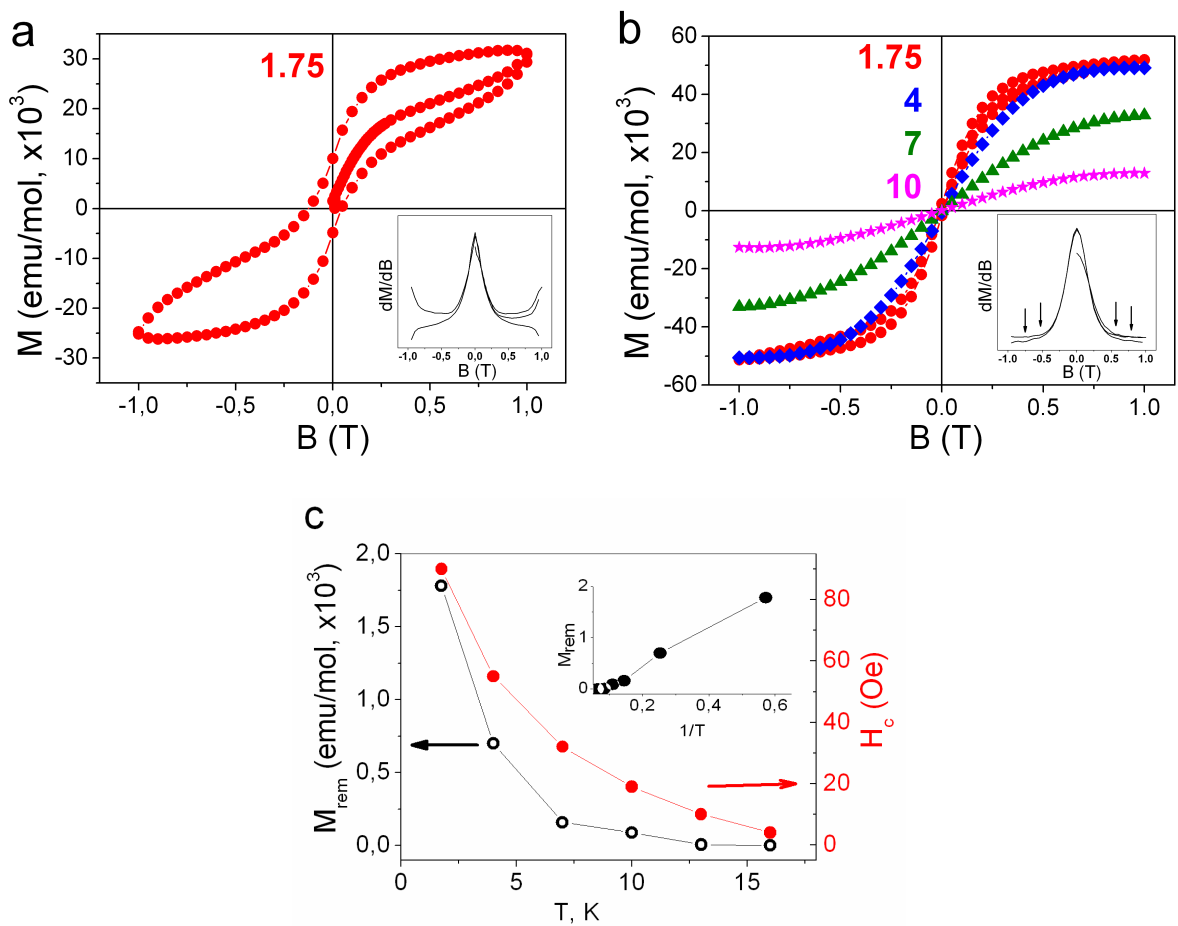

**Fig. 5.** (Color online) Hysteresis loops of complexes **1** (a) and **2** (b); the insets show corresponding derivatives. Remnant magnetization and coercive force as a function of temperature for the complex **2;** inset shows the dependence of remnant magnetization on reciprocal temperature (c).

The substitution of $CH_3$ to $C_6F_5$ leads to even more dramatic changes in the electronic structure and as a result, all exchange interaction switch from antiferromagnetic to ferromagnetic (Fig. 6c), the strongest interaction being between the central atoms. The enhancement in ferromagnetic exchange interaction in the complex with $C_6F_5$ has lead to an interesting effect which we observed experimentally: the molecules are antiferromagnetically coupled to their neighbors. Thus, we have demonstrated both theoretically and experimentally that exchange interactions are the properties of the whole molecule which should be a consequence of a rather strong delocalization of 3d electrons via $p - d$ hybridization.

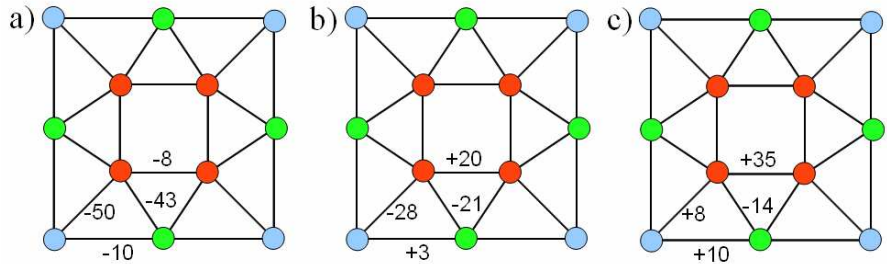

**Fig. 6.** (Color online) A scheme of exchange interactions in $Mn_{12}O_{12}(RCOO)_{16}(H_2O)_4$ for R = $CH_3$ (a), $C_6H_5$ (b), $C_6F_5$ (c). Different types of $Mn^{4+}$ are shown in blue and green, $Mn^{3+}$ in red. Values of exchange interactions reported in Kelvin.

## 4. CONCLUSION

To summarize, we have studied the influence of ligand substitution on magnetic properties of $Mn_{12}$-based SMMs and have shown that it is possible to control the magnetic behavior of a SMM by varying the structure of ligands. In case of benzene derivative, we observed that the magnetostatic parameters as well as the ZFC-FC difference are nonzero at the temperatures exceeding the blocking temperature for the known $Mn_{12}$-derivatives. This behavior, assumingly, is due to π-π stacking between benzoic rings which causes the magnetic core of Mn12 molecule to form a three-dimensional system sufficient to disrupt the pure state of SMM, as the calculations show strong correlation between the magnetic core and the distant ligands. *Ab initio* simulations using the local density approximation with on-site Coulomb repulsion correction (LDA+U) have shown that the ligand variation only negligibly affect the geometry of the inner core of the molecule but instead significantly alters the electronic structure of the SMM: redistribution of the electron density results in the changes of the magnetic exchange interactions between Mn atoms.

## ACKNOWLEDGEMENTS

We thank A.V. Korolev, Yu. Shvachko, D. Starichenko, and D. Baitimirov for the independent proof of the experimental results concerning the field and temperature dependencies of the magnetic moment and for the valuable discussions.